\documentclass[twocolumn,showpacs,preprintnumbers,amsmath,amssymb]{revtex4}
\usepackage{graphicx}% Include figure files
\usepackage{dcolumn}% Align table columns on decimal point
\usepackage{bm}% bold math
\usepackage{tabularx}
\usepackage{rotating}
\usepackage{graphicx}
\usepackage{epsfig}
\usepackage{amsmath,amssymb,bbm}
\usepackage{subfigure}

\begin{document}

\title{A Macromolecule in a Solvent: Adaptive Resolution Molecular
  Dynamics Simulation}
\author{Matej Praprotnik}
\altaffiliation{On leave from the National Institute of Chemistry, Hajdrihova 19,
                 SI-1001 Ljubljana, Slovenia. Electronic Mail: praprot@cmm.ki.si}
\author{Luigi Delle Site}
\author{Kurt Kremer}
\affiliation{%
Max-Planck-Institut f\"ur Polymerforschung, Ackermannweg 10, D-55128 Mainz, Germany
}%

%====================================================================Abstract
\begin{abstract}
We report adaptive resolution molecular dynamics simulations of a
flexible linear polymer in solution. The solvent, i.e., a liquid
of tetrahedral molecules, is represented within a certain radius from the
polymer's center of mass with a high level of
detail, while a lower coarse-grained resolution is used for the
more distant solvent. The high resolution sphere moves with
the polymer and freely exchanges molecules with the low resolution
region through a transition regime. The solvent molecules change
their resolution and number of degrees of freedom on-the-fly.
We show that our approach correctly reproduces the static and
dynamic properties of the polymer chain and surrounding solvent.
\end{abstract}
%===========================================================================
                             % Classification Scheme.
%\keywords{Suggested keywords}%Use showkeys class option if keyword
                              %display desired
\maketitle
%==============================================================================================================INTRO
\section{Introduction}
 The structure of polymers in solution is determined by the
solvent-polymer interaction. In the case of a nonpolar polymer in
an nonpolar solvent, one typically distinguishes three ''types''
of solvent, good, $\Theta$ or marginal, and poor. In the case of a
good solvent the solvent-solvent, solvent-polymer, and polymer-polymer 
interactions effectively result in a situation, where the
chain monomers are preferably surrounded by solvent molecules. As
a result the chains are extended and the size scales as $\langle
R^2 \propto N^{2\nu}\rangle$ with N being the number of monomers and
$\nu \cong 0.6$ in three dimensions. For poor solvent one observes
just the opposite and the chains collapse into a dense globule,
$\langle R^2 \propto N^{2/3}\rangle$. The $\Theta$ regime is where
these two effects compensate and the chains behaves
to a first approximation as a random walk, i.e., $\langle R^2
\propto N\rangle$. In the limit of $N\rightarrow\infty$ the
$\Theta$-point is a tricritical point in the phase diagram. 
As long as the solvent does
not induce special local correlations beyond an unspecific
attraction/repulsion and one is not studying dynamical
properties the collapse of polymers is usually studied with an
implicit solvent. The complicated local interactions are accounted
for by an effective interaction between the chain beads. Studies
of that kind have a long tradition in polymer science and the behavior
is now well understood. Beyond that there
are however many situations, where it becomes difficult or even
questionable, to ignore the local structure of the solvent.
Solvent can play an important role in the functional properties
of macromolecules. For example, dehydration studies of proteins
solvated in water demonstrated that at least a monolayer of water
is needed for full protein functionality\cite{Careri:1999}. The
influence of a macromolecule on the structure and dynamics of the
surrounding solvent is also an important issue. Therefore, a detailed study of interactions of a
macromolecule with a solvent beyond effective coupling parameters
is quite often required for an understanding of the
macromolecule's structure, dynamics, and function. To determine
the interactions of a solvent with a macromolecular solute
chemistry specific interactions on the atomic level of detail have
to be considered. However, the resulting solvating phenomena
manifest themselves at mesoscopic and macroscopic
scales\cite{Das:2005} and in the overall structure of the chains. Due
to large number of degrees of freedom (DOFs) such systems are
difficult to tackle using all-atom computer
simulations\cite{Villa:2005}. Moreover, the vast majority of the
simulation time is typically spent treating the solvent and not
the polymer or protein. A step to bridge the gap between the time
and length scales accessible to simulations that still retain an
atomistic level of detail and the solvating phenomena on
longer time and larger length scales, is given by hybrid
multiscale schemes that concurrently couple different physical
descriptions of the system (see, e.g., Refs.
\cite{Rafii:1998,Broughton:1999,Ahlrichs:1999,Malevanets:2000,
Csanyi:2004,Abrams:2005,Neri:2005, Fabritiis:2006}).

Recently, we have proposed an adaptive resolution molecular
dynamics (MD) scheme (AdResS) that concurrently couples the
atomistic and mesoscopic length scales of a generic
solvent\cite{Praprotnik:2005:4,Praprotnik:2006,Praprotnik:2006:1}.
In the first application we studied a liquid of tetrahedral
molecules where an atomistic region was separated from the mesoscopic
one by a flat or a spherical boundary. The two regimes with
different resolutions freely exchanged molecules while maintaining
the thermodynamical equilibrium in the system. The spatial regions
of different resolutions, however, remained constant during the
course of the simulations. More recently this approach was
extended to the study of water\cite{Praprotnik:2006:2}. In the
present paper, we generalize our approach to the study of a
polymer chain in solution. The chain is surrounded by solvent with
''atomistic'' resolution. When the chain moves around, the sphere
of atomistically resolved solvent molecules moves together with
the center of mass of the chain.  In this way the chain is free to
move around, although the explicit resolution sphere is much
smaller than the overall simulation volume. This enables us to
efficiently treat solvation phenomena, because only the solvent in
the vicinity of a macromolecule is represented with a
sufficiently high level of detail to take the specific
interactions between the solvent and the solute into account. 
Solvent farther away from the solute, where the high
resolution is no longer required, is represented on a more
coarse-grained level. In this work, a macromolecule is represented
by a generic flexible polymer chain\cite{Kremer:1990} embedded in
a solvent of tetrahedral molecules introduced in Refs.
\cite{Praprotnik:2005:4,Praprotnik:2006}. This study represents a
first methodological step towards adaptive resolution MD
simulations of systems of biological relevance, e.g., a
protein in water.

The paper is organized as follows: In section II the dual scale
model of a polymer chain in a liquid is presented. The hybrid
numerical scheme and computational details are given in section
III. The results and discussion are reported in section IV,
followed by a summary and outlook in section V.

\section{Multiscale Model}
\begin{figure}[!ht]
 \centering
\includegraphics[width=8.0cm]{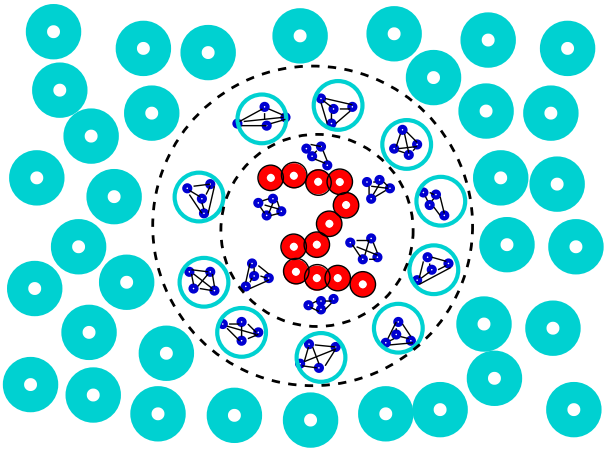}
\caption{(Color online) A schematic plot of a solvated generic
bead-spring polymer. The solvent is modeled on different levels of
detail:  solvent molecules within a certain radius from the
polymer's center of mass are represented with a high (atomistic)
resolution while a lower mesoscopic resolution is used for the
more distant solvent. The high resolution sphere moves with
the polymer's center of mass. The polymer beads are represented
smaller than the solvent molecules for presentation
convenience, for details see text.}\label{Fig.1}
\end{figure}

We study a single generic bead spring polymer solvated in a
molecular liquid as illustrated in figure \ref{Fig.1}.
Solvent molecules within a distance $r_0$ from the polymer's
center of mass are modeled with all 'atomistic' details to
properly describe the specific polymer-solvent interactions. For
the description of the solvent farther away, where the high
resolution is not required, we use a lower resolution. The solvent
molecules then, depending on their distance to the polymer's
center of mass, automatically adapt their resolution on-the-fly.

The model solvent is a liquid of $n$ tetrahedral molecules as
introduced in Refs. \cite{Praprotnik:2005:4,Praprotnik:2006}. The
solvent molecules in the high resolution regime are composed of
four equal atoms with mass $m_0$. Their size $\sigma$ is fixed via
the repulsive Weeks-Chandler-Andersen potential
\begin{multline}
 U_{rep}^{atom}(r_{i\alpha j\beta})=\\\left\{\begin{array}{rc}
4\varepsilon\bigl[\bigl(\frac{\sigma}{r_{i\alpha
      j\beta}}\bigr)^{12}-\bigl(\frac{\sigma}{r_{i\alpha j\beta}}\bigr)^6+\frac{1}{4}\bigr];
& r_{i\alpha j\beta}\le 2^{1/6}\sigma\\
                          0; & r_{i\alpha j\beta}> 2^{1/6}\sigma
                             \end{array}\right.\label{eq.1}
\end{multline}
with the cutoff at $2^{1/6}\sigma$. $\sigma$ and $\varepsilon$ are
the standard Lennard Jones units for lengths and energy
respectively. $r_{i\alpha j\beta}$ is the distance between the
atom $i$ of the molecule $\alpha$ and the atom $j$ of the molecule
$\beta$. The neighboring atoms in a given molecule $\alpha$ are
connected by finite extensible nonlinear elastic (FENE) bonds
\begin{multline}
 U_{bond}^{atom}(r_{i\alpha j\alpha})=\\\left\{\begin{array}{rc}
                  -\frac{1}{2}kR_0^2\ln\bigl[1-\bigl(\frac{r_{i\alpha
 j\alpha}}{R_0}\bigl)^2\bigr]; & r_{i\alpha j\alpha}\le R_0\\
                       \infty; & r_{i\alpha j\alpha}> R_0
                  \end{array}\right.\label{eq.2}
\end{multline}
with divergence length  $R_0=1.5\sigma$ and stiffness
$k=30\varepsilon/\sigma^2$, so that the average bond length is
approximately $0.97\sigma$ for $k_BT=\varepsilon$, where $T$ is the
temperature of the system and $k_B$ is Boltzmann's constant. For
the coarse-grained solvent model in the low resolution regime we
use one-site spherical molecules interacting via an effective pair
potential\cite{Praprotnik:2006}, which was derived such that the
statistical properties, i.e., the center of mass radial
distribution function and pressure, of the high resolution liquid
are accurately reproduced. This is also needed for the present
study, since the motion of the high resolution sphere should not
be linked to strong rearrangements in the liquid. The high and low
resolution freely exchange molecules through a transition regime
containing hybrid molecules (see figure \ref{Fig.1}), where the
molecules with no extra equilibration adapt their
resolution and change the number of DOFs
accordingly\cite{Praprotnik:2005:4,Praprotnik:2006,Praprotnik:2006:1}.

The polymer is modeled as a standard bead-spring polymer
chain\cite{Kremer:1990}. It contains $N$ monomers, which represent
chemical repeat units, usually comprising several atoms. The
interactions between monomers (beads) are defined using Eqs.
(\ref{eq.1}) and (\ref{eq.2}) with the rescaled values
$\sigma_B=1.8\sigma$, $R_{0_B}=R_0\sigma_B/\sigma=1.5\sigma_B$,
and $k_B=k\sigma^2/\sigma_B^2=30\varepsilon/\sigma_B^2$, such that
the size of the polymer bead $\sigma_B$ is approximately the same
as the size of the solvent molecule\cite{Praprotnik:2006}. The
average bond length between beads is rescaled accordingly. The
bead mass is also increased $m_B=5m_0$ to make them behave more
like Brownian particles. Standard Lorentz-Berthelot mixing
rules\cite{Allen:1987} are used for the interaction between
monomers and the 'atoms' of the solvent molecules.

\section{Numerical Scheme and Computational Details}
To smoothly couple the regimes of high and low level of detail of
the description of the solvent molecules, we apply the recently
introduced AdResS scheme\cite{Praprotnik:2005:4}. There the
molecules can freely move  between the regimes, they are in
equilibrium with each other with no barrier in between.  The
transition is governed by a weighting function $w(r)\in[0,1]$ that
interpolates the molecular interaction forces between the two
regimes, and assigns the identity of the solvent molecule. We
resort here to the weighting function defined in Ref.
\cite{Praprotnik:2006}:
\begin{eqnarray}
 w(r)=\left\{\begin{array}{rc}
                                                   1; & r_0 > r\ge 0\\
                                                   0; & r\ge r_0+d\\
                       \cos^2[\frac{\pi}{2d}(r-r_0)]; & r_0+d > r\ge r_0
            \end{array}\right.\label{eq.3}
\end{eqnarray}
where $r_0$ is the radius of the high resolution region and $d$
the interface region width, cf. \ref{Fig.1}. The radius $r_0$ must
be chosen sufficiently large so that the whole polymer always
stays within the high resolution solvent regime. $w(r)$ is defined
in such a way that $w=1$ corresponds to the high resolution, $w=0$
to the low resolution, and values $0<w<1$ to the transition regime,
respectively. This leads to intermolecular force acting between
centers of mass of solvent molecules $\alpha$ and $\beta$:
\begin{multline}
{\bf F}_{\alpha\beta}=w(|{\bf R}_\alpha-{\bf R}|)w(|{\bf R}_\beta-{\bf
 R}|){\bf
 F}_{\alpha\beta}^{ex}\\+[1-w(|{\bf R}_\alpha-{\bf R}|)w(|{\bf R}_\beta-{\bf
 R}|)]{\bf
 F}_{\alpha\beta}^{cg}.\label{eq.4}
\end{multline}
${\bf F}_{\alpha \beta}$ is the total intermolecular force acting
between centers of mass of the solvent molecules $\alpha$ and
$\beta$. ${\bf F}_{\alpha\beta}^{ex}$ is the sum of all pair
'atom' interactions between explicit tetrahedral 'atoms' of the
solvent molecule $\alpha$ and explicit tetrahedral 'atoms' of the
solvent molecule $\beta$, ${\bf F}_{\alpha\beta}^{cg}$ is the
effective pair force between the two solvent molecules, and ${\bf
R}_\alpha$, ${\bf R}_\beta$, and ${\bf R}$ are the centers of mass
of the molecules $\alpha$, $\beta$ and the polymer, respectively.
Note that one has to interpolate the forces and not the
interaction potentials in Eq. (\ref{eq.4}) if the Newton's Third
Law is to be satisfied~\cite{Praprotnik:2006:1}. To suppress the
unphysical density and pressure fluctuations emerging as artifacts
of the scheme given in Eq. (\ref{eq.4}) within the transition zone
we employ an interface pressure correction\cite{Praprotnik:2006}.
The latter involves a reparametrization of the effective
potential in the system composed of exclusively hybrid molecules with $w=1/2$.
Each time a solvent molecule crosses a boundary between the
different regimes it gains or looses on-the-fly (depending on
whether it leaves or enters the coarse-grained region) its
equilibrated rotational and vibrational DOFs while retaining its
linear momentum\cite{Praprotnik:2006:1, Praprotnik:2005,
Praprotnik:2005:1, Praprotnik:2005:2}. This change in resolution
requires to supply or remove ''latent heat'' and thus must be
employed together with a thermostat that couples locally to the
particle motion~\cite{Praprotnik:2005:4,Praprotnik:2006:1}. This
is achieved by coupling the particle motion to the Dissipative
Particle Dynamics (DPD)
thermostat\cite{Soddemann:2003}. This bears the additional
advantage of preserving momentum conservation and correct
reproduction of hydrodynamics in our $nVT$ MD simulations. Because
of the freely moving polymer chain and solvent molecules, the
above scheme requires the center of the high resolution sphere to
move with the polymer but slowly compared to the
surrounding solvent molecules, so that they at the boundary
between different regimes have enough time to adapt to the
resolution change. The validity condition for our approach thus
requires $D_{polymer} \ll D_{solvent}$, where $D_{polymer}$ and
$D_{solvent}$ the corresponding diffusion constants. This
condition is trivially fulfilled in polymeric solutions and thus
also in our simulations (see the next section).

We conducted all MD simulations using the ESPResSo
package\cite{Espresso:2005}. We integrated Newtons equations of
motion by a standard velocity Verlet algorithm with a time step
$\Delta t=0.005 \tau$ and coupled the motion of the particles to a
DPD theromstat\cite{Soddemann:2003} with the temperature set to
$T=\varepsilon/k_B$. The DPD friction constant
$\zeta=0.5\tau^{-1}$, where
$\tau=(\varepsilon/m_0\sigma^2)^{-1/2}$, and the DPD cutoff radius
was set equal to the cuttoff radius of the effective pair
interaction between solvent molecules, i.e.,
$3.5\sigma$\cite{Praprotnik:2006}. The width of the transition
regime is $2.5\sigma$\cite{Praprotnik:2006}. Periodic boundary conditions and the minimum image
convention\cite{Allen:1987} were employed. After equilibration,
trajectories of $5000\tau$ were obtained, with configurations
stored every $5\tau$. These production runs were performed with a
$10^{9}\varepsilon/\sigma$ force capping to prevent possible force
singularities that could emerge due to overlaps with the
neighboring molecules when a given molecule enters the transition
layer from the coarse-grained side \cite{Praprotnik:2005:4}. The
temperature was calculated using the fractional analog of the
equipartition theorem:
\begin{equation}
 \left<K_\alpha\right>=\frac{\alpha k_BT}2,
\label{eq4}
\end{equation}
where $\left<K_\alpha\right>$ is the average kinetic energy per
fractional quadratic DOF with the weight
$w(r)=\alpha$\cite{Praprotnik:2006:1}. Via Eq. (\ref{eq4}) the
temperature is also rigorously defined in the transition regime in
which the vibrational and rotational DOFs are partially 'switched
on/off'. The molecular number density of the solvent is
$\rho=0.175/\sigma^3$, which corresponds to a typical high density
Lennard-Jones liquid\cite{Praprotnik:2006}. We considered three
different system sizes with corresponding cubic box sizes:
$L=25.0\sigma, 30.6\sigma, 34.2\sigma$. The reduced Lennard-Jones
units\cite{Allen:1987} are used in the remainder of the paper.

\section{Results and Discussion}

To validate the AdResS approach for the present polymer solvent
system, we carried out the analysis  of the structural and dynamic
properties of a polymer chain in the hybrid multiscale solvent
compared to the corresponding fully explicit system where all
solvent molecules are modeled with a high level of detail, i.e.,
as a tetrahedral molecules.

\subsection{Statics of the Polymer Chain and Solvent}
First, we focus on the explicit (\emph{ex}) systems where the
solvent is modeled with the high resolution all over the
simulation box. These results are considered as the reference to
check how well AdResS produces the same physics as the all-atom MD
simulation. The reference average thermodynamic properties of the
corresponding \emph{ex} systems (polymer+explicitely resolved
solvent) are listed in table \ref{Tab.1a}.
\begin{table}[ht!]
   \centering
   \begin{tabular}{||c|c|c|c||}
    \hline\hline
      $N$                             &    $10$           &  $20$            & $30$    \\
       $L$                            &    $25.0$  & $30.6$  & $34.2$\\
 \hline
      $<p>$                           &    $2.01\pm 0.04$ &  $2.01\pm 0.03$  &  $2.02\pm 0.01$\\
      $<T>$                           &    $1.0\pm 0.01$  &  $1.0\pm 0.01$   &  $1.0\pm 0.01$\\
      $<T_{polymer}>$                 &    $1.0\pm 0.5$   &  $1.0\pm 0.3$    &  $1.0\pm 0.2$\\
      $<T_{solvent}>$                 &    $1.0\pm 0.01$  &  $1.0\pm 0.01$   &  $1.0\pm 0.01$\\
     \hline\hline
   \end{tabular}
   \caption{Thermodynamic properties of the fully explicit systems (\emph{ex}) of different chain lengths
   $N$ and box sizes $L$:  average total pressure
   $\left<p\right>$,  average total temperature of the system
   $\left<T\right>$, average temperature of the polymer
   $\left<T_{polymer}\right>$, and average temperature of the solvent
   $\left<T_{solvent}\right>$.} \label{Tab.1a}
  \end{table}

%%%%%%%%%%%%%%%%%%%%%%%%%%%%%%%%%%%%%%%%%%%%%%%%%%%%%%%%%%%%%%%%%%%%%%%%%%%%%%%%%%%%%%%%%%%%%%%%%%%%

\begin{figure}[ht!]
\centering
\subfigure[]{\includegraphics[width=7.2cm]{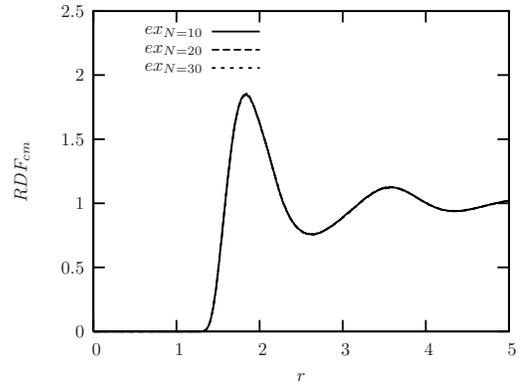}}
\subfigure[]{\includegraphics[width=7.2cm]{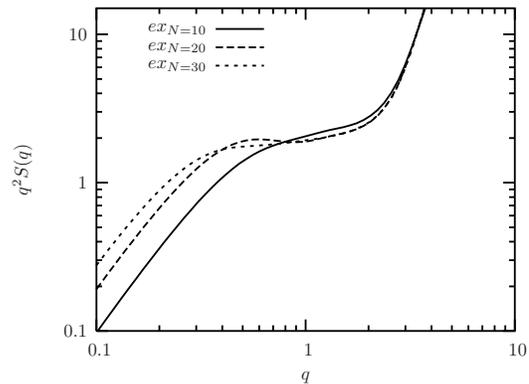}}
\subfigure[]{\includegraphics[width=7.2cm]{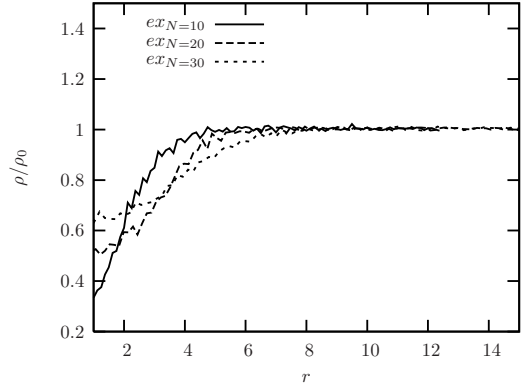}}

\caption{ (a) The solvent center of mass RDF for three different
polymer lengths (N=10,20,30), (b) the static structure factor of
the polymer in the Kratky representation, and (c) the solvent
density around the center of mass of the chains, which illustrates
the so called correlation hole.
 }\label{Fig.2}
\end{figure}

The static properties of the solvent are characterized by the
solvent radial center of mass distribution (RDF) function depicted
in figures \ref{Fig.2} (a). This distribution function is within
the thickness of the lines the same for all systems studied,
including the hybrid ones. This is to be expected from our
previous studies and the fact, that the polymer fraction of volume
is very small compared to that of the solvent.

The statistical properties of polymers are conveniently described
by a number of quantities, namely the radius of gyration
\begin{equation}
 \left<R_G^2\right>=\frac{1}{N}\sum_i \left<({\bf r}_i-{\bf R})^2\right>,
\end{equation}
where ${\bf r}_i$ is the position vector of the $i$th monomer and
${\bf R}=N^{-1}\sum_i{\bf r}_i$ is the polymer's center of
mass, the end-to-end distance
\begin{equation}
 \left<R_E^2\right>=\left<({\bf r}_N-{\bf r}_1)^2\right>,
\end{equation}
and the hydrodynamic radius
\begin{equation}
\left<\frac{1}{R_H}\right>=\frac{1}{N^2}\sum_{i\ne j}\left<\frac{1}{r_{ij}}\right>,\label{eq.rh}
\end{equation}
where $r_{ij}=|{\bf r}_i-{\bf r}_j|$\cite{Praprotnik:2006:3}.

$\left<R_G^2\right>$ and $\left<R_E^2\right>$ scale as
\begin{equation}
\left<R_G^2\right>\propto\left<R_E^2\right>\propto N^{2\nu}
\end{equation}
with the number of monomers $N$ where $\nu=0.5$ in $\theta$
solvent and $\nu\approx 0.588$ in good solvent
conditions\cite{Gennes:1979,Doi:1986,Sokal:1995} with rather small
finite size corrections, while the hydrodynamic radius is known to
show significant deviations from asymptotic behavior up to very
long chains\cite{Dunweg:1993}.

The single-chain static structure factor $S(q)$
\begin{equation}
 S(q)=\frac{1}{N}\left<\sum_{ij}\exp(i{\bf
   q}\cdot({\bf r}_i-{\bf r}_j))\right>
\end{equation}
probes the self similar structure within the scaling regime and
 thus provides an accurate way to determine $\nu$. $S(q)$ scales as
\begin{equation}
 S(q)\propto q^{-1/\nu}\rightarrow q^2S(q)\propto q^{2-1/\nu}\label{eq.Sq}
\end{equation}
in the regime $R_G^{-1}\ll q\ll b^{-1}$, where $b$ is the typical
bond length. By fitting a power law to the computed $q^2S(q)$
plotted in figure \ref{Fig.2} (b) we obtained the values for
$\nu$ reported in table \ref{Tab.1}. Table \ref{Tab.1} summarizes the values of all quantities
defined above, which characterize the static properties of the polymer
chain. The calculations were performed for $N=10,20,30$.
  \begin{table}[ht!]
   \centering
   \begin{tabular}{||c|c|c|c||}
    \hline\hline
      $N$                             &    $10$           &  $20$            & $30$    \\
      $L$                             &    $25.0$         &  $30.6$          &  $34.2$ \\
    \hline
      $\left<\Delta r_{max}\right>$   &    $4.2\pm 0.8$   &  $5.7\pm 1.0$    &  $8.1\pm 1.4$\\
      $R_G=\left<R_G^2\right>^{1/2}$  &    $2.7\pm 0.5$   &  $3.8\pm 0.6$    &  $5.0\pm 0.8$\\
      $R_E=\left<R_E^2\right>^{1/2}$  &    $6.7\pm 2.0$   &  $8.6\pm 3$    &  $12\pm 3$\\
      $R_H=\left<R_H^{-1}\right>^{-1}$&    $3.3\pm 0.3$   &  $4.0\pm 0.3$    &  $4.7\pm 0.4$\\
      $\nu$                           &    $0.63$         &  $0.54$  &  $0.57$\\
      $2/z$                           &    $0.59$         &  $0.71$  &  $0.67$\\
      $\tau=R_G^2/(6D)$               &      $152$        &  $481$           & $1390$\\
     \hline\hline
   \end{tabular}
   \caption{Summary of some polymer (embedded in the explicitely
   resolved \emph{ex} solvent) properties: number of polymer beads $N$, size of the
   simulation box $L$, average maximal
   distance of a monomer from the polymer's center of mass $\left<\Delta
r_{max}\right>$, radius of gyration $R_G$, end-to-end distance
$R_E$, hydrodynamic radius $R_H$, the static exponent $\nu$,  the
exponent $2/z$, where $z$ is the dynamic exponent, and the longest
relaxation time $\tau$ calculated using data from table
\ref{Tab.4}. The error bar for the exponents $\nu$ and $2/z$ is roughly $10\%$.}\label{Tab.1}
  \end{table}

Another property, which directly reveals the fractal structure of
the chains is the correlation hole, which is shown in figure \ref {Fig.2}
(c). It directly shows, to which distance from the center of mass
of the chains, the solvent density is perturbed by the chain
beads. For the later application of the hybrid scheme it is
important to define the explicit solvent regime large enough in
order to cover the correlation hole completely.

The values of $\nu$ actually differ slightly from the asymptotical
value for the good solvent due to the finite chain lengths.
Nevertheless, the agreement improves with the increasing $N$, as
expected.

Let us now turn our attention to the hybrid solvent studied by MD
simulation using AdResS.

\begin{figure}[!ht]
 \centering
\includegraphics[width=7.2cm]{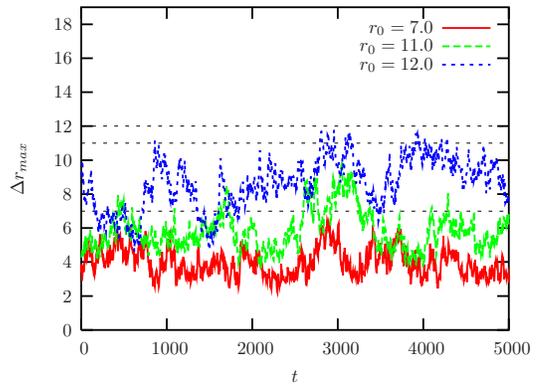}
\caption{(Color online) Time evolution of the maximal monomer
distance from the polymer's center of mass,  $\Delta
r_{max}$, for polymers with $N=10$ beads and the radius of
the high resolution regime $r_0=7.0$ (red line), $N=20$ and
$r_0=11.0$ (green line), and $N=30$ and $r_0=12.0$ (blue
line).}\label{Fig.3}
\end{figure}

To assure that the polymer is surrounded only by the explicitely
resolved molecules, we determine first the maximal monomer
distance from the polymer's center of mass, $\Delta
r_{max}$, as shown in figure \ref {Fig.3} for all chain lengths
studied.
As shown $\Delta r_{max}$ always stays within the
high resolution regime.

\begin{figure}[!ht]
\centering
\subfigure[]{\includegraphics[width=7.2cm]{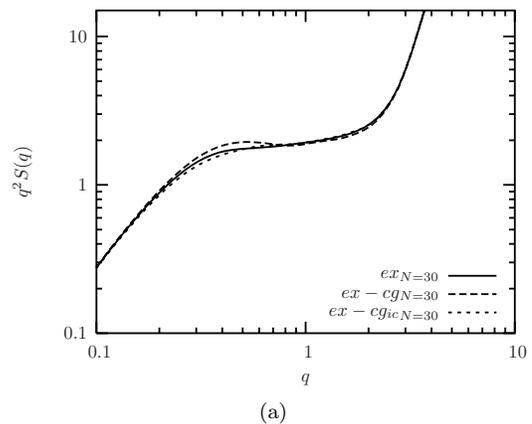}}
\subfigure[]{\includegraphics[width=7.2cm]{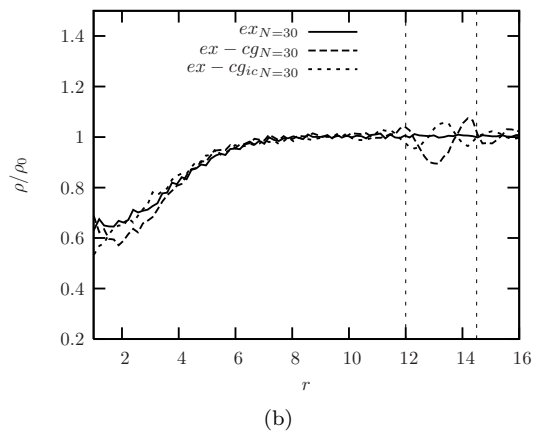}}
\caption{(a) The the static structure factor of the polymer with
            $N=30$ in the Kratky representation for all three cases studied: the
            fully explicite, the AdResS scheme with and without the
            interface pressure correction. (b) The correlation hole for the same systems
            as in (a).
 }\label{Fig.4}
\end{figure}

Because for $N=30$ $\Delta r_{max}$ gets rather close
to $r_0$, we checked the static polymer properties for that case
again. In figure \ref{Fig.4} we compare the all explicit
simulation to the two hybrid simulation schemes (with
(\emph{ex-cg}$_{ic}$) and without (\emph{ex-cg})
the pressure correction\cite{Praprotnik:2006} in the transition regime) for the chain
form factor and the correlation hole. The agreement is excellent,
showing that the proposed scheme should at least be capable of
properly reproducing the conformational statistics of the embedded
polymer in solution.

This is first checked by comparing the average thermodynamic
properties as given in table \ref{Tab.2a} for the hybrid
\emph{ex-cg} and \emph{ex-cg}$_{ic}$ systems (polymer+hybrid
solvent). While the temperatures are identical the pressure
correction in the interface layer reduces the pressure slightly,
so that the hybrid system now also there agrees quite well with the
all explicit simulation.
\begin{table}[ht!]
   \centering
   \begin{tabular}{||c|c|c|c||}
     \hline\hline
      $N$                             &    $10$           &  $20$            & $30$    \\
       $L$                            &    $25.0$  & $30.6$  & $34.2$\\
 \hline
     $<p>_{ex-cg}$                    &    $2.03\pm 0.02$ &  $2.04\pm 0.01$  &  $2.04\pm 0.03$\\
     $<p>_{ex-cg_{ic}}$               &    $2.01\pm 0.01$ &  $2.01\pm 0.01$  &  $2.01\pm 0.01$\\
       $<T>$                           &    $1.0\pm 0.02$  &  $1.0\pm 0.01$   &  $1.0\pm 0.01$\\
      $<T_{polymer}>$                 &    $1.0\pm 0.5$   &  $1.0\pm 0.3$    &  $1.0\pm 0.2$\\
      $<T_{solvent}>$                 &    $1.0\pm 0.02$  &  $1.0\pm 0.01$   &  $1.0\pm 0.01$\\
     \hline\hline
   \end{tabular}
   \caption{Thermodynamic properties of systems with the polymer
   solvated in the hybrid \emph{ex-cg} solvent and the hybrid \emph{ex-cg}$_{ic}$
   solvent:  average total pressure
   $\left<p\right>$,  average total temperature of the system
   $\left<T\right>$, average temperature of the polymer
   $\left<T_{polymer}\right>$, and average temperature of the solvent
   $\left<T_{solvent}\right>$. For the temperatures the results cannot be distinguished.} \label{Tab.2a}
  \end{table}
The agreement with the reference values from table \ref{Tab.1a} is
very good. This is in line with the general static properties of
the polymers, which are given in  \ref{Tab.2} and  \ref{Tab.3}, and
compare very well to the data from table \ref{Tab.1}.
\begin{table}[ht!]
   \centering
   \begin{tabular}{||c|c|c|c||}
    \hline\hline
      $N$                             &    $10$           &  $20$            &  $30$    \\
      $L$                             &    $25.0$         &  $30.6$          &  $34.2$ \\
    \hline
      $r_0$                           &    $7.0$          &  $11.0$          &  $12.0$ \\
      $\left<\Delta r_{max}\right>$   &    $4.2\pm 0.8$   &  $6.6\pm 1.2$    &  $7.7\pm 1.3$\\
      $R_G=\left<R_G^2\right>^{1/2}$  &    $2.7\pm 0.4$   &  $4.0\pm 0.6$    &  $4.6\pm 0.7$\\
      $R_E=\left<R_E^2\right>^{1/2}$  &    $6.7\pm 2.0$   &  $10.4\pm 2.8$   &  $10.8\pm 2.5$\\
      $R_H=\left<R_H^{-1}\right>^{-1}$&    $3.3\pm 0.3$   &  $4.0\pm  0.3$   &  $4.5\pm 0.4$\\
      $\nu$                           & $0.63$            &  $0.58$  &  $0.54$\\
      $2/z$                           & $0.56$            &  $0.69$  &  $0.62$\\
      $\tau=R_G^2/(6D)$               &      $122$        &  $381$           &  $882$\\
    \hline\hline
   \end{tabular}
   \caption{Summary of some polymer (embedded in the hybrid \emph{ex-cg}
   solvent) properties: number of polymer beads $N$, size of the
   simulation box $L$, radius of the high resolution regime $r_0$, average maximal
   distance of a monomer from the polymer's center of mass
   $\left<\Delta r_{max}\right>$, radius of gyration $R_G$, end-to-end
   distance $R_E$, hydrodynamic radius $R_H$, the static exponent
   $\nu$, the exponent $2/z$, where $z$ is the dynamic exponent, and the
   longest relaxation time $\tau$ calculated using data from table
   \ref{Tab.4}. The error bar for the exponents $\nu$ and $2/z$ is roughly $10\%$.} \label{Tab.2}
  \end{table}

\begin{table}[ht!]
   \centering
   \begin{tabular}{||c|c|c|c||}
    \hline\hline
      $N$                             &    $10$           &  $20$            &  $30$    \\
      $L$                             &    $25.0$         &  $30.6$          &  $34.2$ \\
    \hline 
      $r_0$                           &    $7.0$          &  $11.0$          &  $12.0$ \\
      $<\Delta r_{max}>$              &    $4.0\pm 0.8$   &  $5.9\pm 1.2$    &  $8.6\pm 1.5$\\
      $R_G=\left<R_G^2\right>^{1/2}$  &    $2.7\pm 0.5$   &  $3.9\pm 0.7$    &  $5.2\pm 0.8$\\
      $R_E=\left<R_E^2\right>^{1/2}$  &    $6.6\pm 2.1$   &  $9.4\pm 3.0$    &  $13.3\pm 3.3$\\
      $R_H=\left<R_H^{-1}\right>^{-1}$&    $3.2\pm 0.3$   &  $4.0\pm 0.4$    &  $4.8\pm 0.4$\\
      $\nu$                           & $0.59$            &  $0.55$  &  $0.57$\\
      $2/z$                           & $0.69$            &  $0.71$  &  $0.77$\\
      $\tau=R_G^2/(6D)$               &      $152$        &  $362$           &  $1127$\\
   \hline\hline
   \end{tabular}
   \caption{Same data as in table \ref{Tab.2}, but now for the hybrid \emph{ex-cg}$_{ic}$, where
  interface pressure correction is applied.} \label{Tab.3}
  \end{table}

From the presented results we can conclude that AdResS faithfully
reproduces the reference statics obtained from the simulations
with a polymer embedded in the explicitely resolved solvent.

\subsection{Dynamics of the Polymer Chain and Solvent}

While the conformational properties of the polymer in solution are
well understood and properly described by the adaptive resolution
approach, the situation for the dynamics is much less clear. By
changing the degrees of freedom not only the structure but also
the dynamical properties are altered, however in a way which is less
understood. It is also not a priori clear, whether an approach,
which produces a precise coarse graining for structural properties,
does this for dynamical properties as well. In a recent
study of small additive molecules to a polymer melt it was shown,
that while the length scaling is identical, the time scaling can
be different\cite{Harmandaris:2007}. In the present situation 
the influence of the transition regime poses additional difficulties. 

In order to
determine the dynamical properties of the solvent and solute we
calculated the respective diffusion coefficients. The diffusion
coefficient of a species is computed from the center of mass
displacements using the Einstein relation
 \begin{equation}
  D=\frac{1}{6}\lim_{t\rightarrow\infty}\frac{\langle|{\bf
  R}_i(t)-{\bf
  R}_i(0)|^2\rangle}{t}=\frac{1}{6}\lim_{t\rightarrow\infty}\frac{\langle\Delta
  R^2\rangle}{t},
 \end{equation}
where ${\bf R}_i(t)$ is the center-of-mass position of the
molecule $i$ (which can be either a solvent or a solute molecule)
at time $t$ and averaging is performed over all choices of time
origin and, in the case of solvent, over all solvent molecules.

\begin{figure}[!ht]
\centering
\subfigure[]{\includegraphics[width=7.2cm]{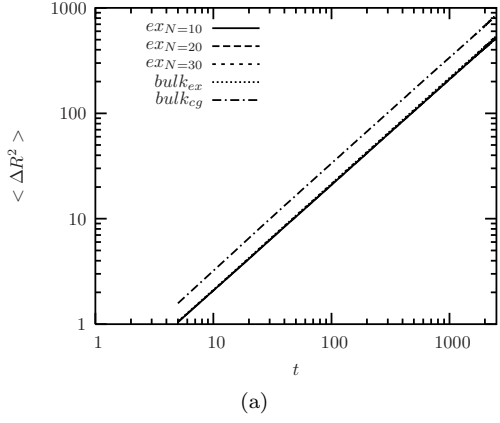}}
\caption{Log-log plot of the time dependence of the mean square
 displacement of the solvent molecule's center of mass in the.
 time interval [0,5000]: explicitely resolved
 (bulk$_{ex}$) and coarse-grained (bulk$_{cg}$) solvents without solvated
 polymer and the explicitely resolved solvent for the systems with three different
 lengths of the solvated polymer
 (N=10,20,30).
 }\label{Fig.5}
\end{figure}

Figure \ref{Fig.5} shows this for the solvent molecules' centers
of mass as a function of time for the different systems indicated.
All the curves in figure \ref{Fig.5}, except the one for the
coarse-grained solvent coincide. Thus the effect of the polymer on
the diffusivity of the solvent molecules is negligible. In other
words, the dilution is strong enough that the polymer effect on
the solvent dynamics is very small. The coarse grained solvent
molecules however move faster than the explicit ones. This is a
consequence of the reduced number of DOFs causing a time scale
difference in the dynamics of the coarse-grained
system\cite{Praprotnik:2005:4, Praprotnik:2006}. While this can be
very advantageous in some cases, one can also adjust $D$ by an
increased background friction in the DPD
thermostat\cite{Kremer:1990,Izvekov:2006}.
The diffusion coefficient $D$ of the solvent was obtained by
fitting a linear function to the curves depicted in figure
\ref{Fig.5} (a) and the obtained values are $D_{bulk_{ex}}=0.036$
and $D_{bulk_{cg}}=0.057$ for the explicit and coarse-grained
solvent, respectively. The question to ask here is, to what extend
does this have any influence on the dynamics of the embedded
polymer.

\begin{figure}[!ht]
\centering
\subfigure[]{\includegraphics[width=7.2cm]{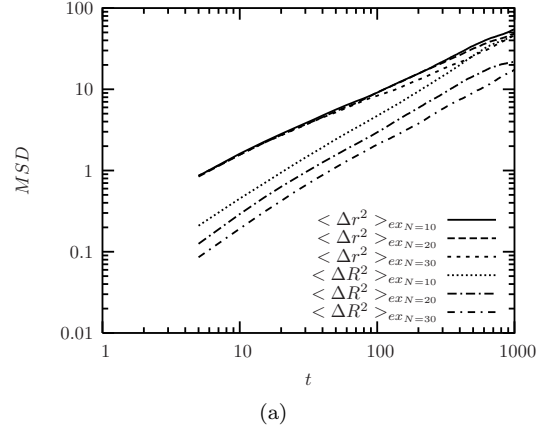}}
\subfigure[]{\includegraphics[width=7.2cm]{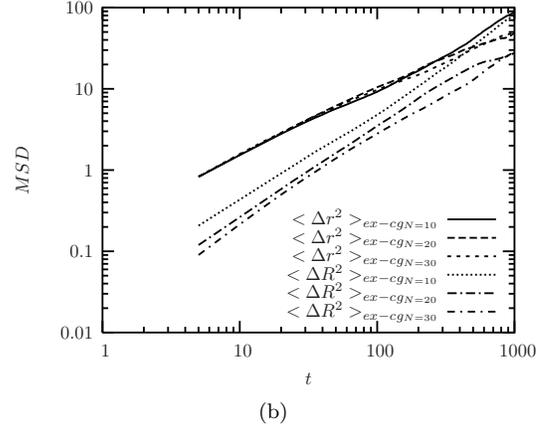}}
\subfigure[]{\includegraphics[width=7.2cm]{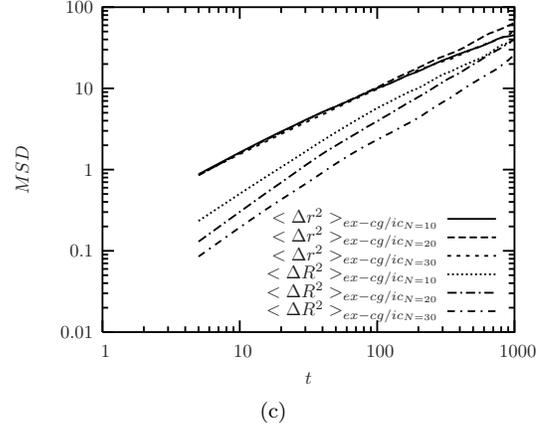}}
\caption{Log-log plot of the time dependence of the mean square
displacement of a single monomer (considered are only monomers
near the chain's center of mass) for polymers and their centers of
mass with $N=10,20,30$ solvated in the \emph{ex} solvent (a), the
hybrid \emph{ex-cg} solvent (b) and the \emph{ex-cg}$_{ic}$
solvent (c) as indicated.
 }\label{Fig.6}
\end{figure}

Within the Zimm model\cite{Doi:1986} for polymer chain dynamics,
which is known to describe the scaling of the dynamics in dilute
solutions of polymers rather well and which takes into account the
hydrodynamic interactions, the polymer diffusion coefficient
scales as
 \begin{equation}
  D\propto N^{-\nu}\propto R_H^{-1}\propto R_G^{-1}.
 \end{equation}
The longest relaxation time $\tau=R_G^2/(6D)$, i.e., the Zimm time
$\tau_Z \propto R_G^3 = R_G^z$, is the time the chain needs to
move its own size. $z=3$ is the dynamic exponent. Note that the
motion of inner monomers within the appropriate scaling regime
should be independent of $N$. For the mean square displacements of
the monomers a scaling analysis immediately yields for the mean
square displacement of a monomer $i$,
 \begin{equation}\langle\Delta r^2\rangle=\langle({\bf r}_i(t)-{\bf
r}_i(0))^2\rangle \propto t^{2/z}=t^{2/3},
 \end{equation}
for distances significantly larger than the bond length and
smaller than $\langle R^2 \rangle$, i.e., times smaller than
$\tau_Z$. For the center of mass of the chains a diffusive
behavior for the mean square displacement $\langle\Delta R^2 (t)
\rangle$ is always observed. Although the chains are relatively short, at
least for $N=30$ one expects a behavior relatively close to the
above mentioned idealized scheme\cite{Dunweg:1993}. Figure
\ref{Fig.6} shows $\langle\Delta r^2 (t) \rangle$ and
$\langle\Delta R^2 (t) \rangle$ for polymer chains with
$N=10,20,30$ embedded in the different solvent scenarios studied.

In all cases the observed exponents for $\langle\Delta r^2 (t)
\rangle$ and $\langle\Delta R^2 (t) \rangle$ are close to $0.7\pm
0.05$ and $1\pm 0.05$ respectively. Also the amplitudes of the
displacements of the inner beads are almost the same for all chain
lengths studied. This is in good agreement with earlier studies on
different generic polymer models in a explicit solvent as well as
studies of chains in a hybrid lattice Boltzmann
solvent\cite{Dunweg:1993,Ahlrichs:1999}. This is
to be expected since we preserve the hydrodynamic interactions by
employing the DPD thermostat in our simulations. This suggests
that our hybrid scheme is also applicable to study dynamic
properties of a polymer in a solution. Small deviations however
occur in the diffusion constant itself. The diffusion constants
for the polymers  with $N=10,20,30$ in the hybrid \emph{ex-cg} and
\emph{ex-cg}$_{ic}$ solvents were obtained by fitting the straight
curve to the polymer's center of mass mean square displacement
presented in figure \ref{Fig.6}. The fit yields data listed in
table \ref{Tab.4}. The corresponding static and dynamic exponents
and the longest relaxation times are given in tables \ref{Tab.2}
and \ref{Tab.3}.

 \begin{table}[ht!]
   \centering
   \begin{tabular}{||c|c|c|c||}
    \hline\hline
      $N$                    &    $10$           &  $20$            &  $30$    \\
    \hline
      D(\emph{ex})           &    $0.008$        &  $0.005$         &  $0.003$ \\
      D(\emph{ex-cg})        &    $0.009$        &  $0.006$         &  $0.0045$ \\
      D(\emph{ex-cg}$_{ic}$) &    $0.0085$        &  $0.006$         &  $0.0035$ \\
    \hline\hline
   \end{tabular}
   \caption{Diffusion constant of the polymer chain embedded in three
different solvents: explicitely resolved \emph{ex}, hybrid
\emph{ex-cg}, and hybrid \emph{ex-cg}$_{ic}$. Though the
statistics of the data is rather poor we can estimate the error
bar roughly to $10 - 15 \%$. For comparison, the diffusion
coefficients of the explicit and coarse-grained solvents are
$D_{bulk_{ex}}=0.036$ and $D_{bulk_{cg}}=0.057$, respectively with
an effect of the polymers too small to determine here. Hence
$D_{polymer} \ll D_{solvent}$.} \label{Tab.4}
  \end{table}

While the ratio of the diffusion constants for different chain
lengths roughly follow the expected scaling, even though it cannot
hold precisely due to the different box sizes, we here observe a
tendency to a weakly accelerated diffusion in the hybrid regime.
This is most evident for the hybrid \emph{ex-cg} case. Two different
aspects might play a role here. First the viscosity in the coarse
grained outer regime is smaller, which must have an effect on the
diffusion. Second, the small pressure and density fluctuations in
the transition regime might contribute to the effect as well.

Although this is only a very first and incomplete test, it shows
that within the AdResS scheme essential aspects of the dynamical
properties of the embedded polymer chain are reasonably well
reproduced.

\section{Summary and Outlook}
In this paper we presented a hybrid multiscale MD simulation of a
generic macromolecule in a solvent using the recently proposed
AdResS method. The solvent surrounding the macromolecule is
represented with a sufficiently high level of detail so that the
specific interactions between the solvent and the solute are correctly
taken into account. The solvent farther away from the
macromolecule, where the high resolution is not needed, is
represented on a coarse-grained level. The high and low resolution
regimes freely exchange solvent molecules, which change 
their resolution accordingly. To correctly simulate momentum
transport through the solvent, we use the DPD thermostat. The simulation results show
that AdResS accurately reproduces the thermodynamic and structural
properties of the system. The presented methodology is an
extension of AdResS to simulations of a solvation cavity, and
represents a first step towards the treatment of more realistic
systems such as biomacromolecules embedded in water. Work along
these lines is already under way\cite{Praprotnik:2006:2}.

\section*{\small ACKNOWLEDGMENTS}
We thank Thomas Vettorel, Vagelis Harmandaris, Benedict Reynolds, and Burkhard D{\"
u}nweg for useful discussions. This work is supported in part by
the Volkswagen foundation. One of the authors (M.~P.) acknowledges
the financial support from the state budget by the Slovenian
research Agency under grant No. P1-0002.

%============================================================================================================

%====================================================================
%\bibliographystyle{physics}

%\bibliography{bibliography}

\begin{thebibliography}{10}
\bibitem{Praprotnik:2006:3}
Note that the definition of $R_H$ given in Eq. (\ref{eq.rh}) is only correct
  for an infinite simulation box. For a finite box we have to also consider the
  hydrodynamic interactions with the periodic images. The value of in such a
  way corrected $R_H$ deviates from the value obtained using the formula
  (\ref{eq.rh})\cite{Dunweg:1993,Ahlrichs:1999}. However, since our aim in the
  present work is only to compare the properties of the polymer chain from the
  hybrid simulation with the corresponding ones from the all-atom simulation we
  here do not take the finite size correction into account and use the formula
  (\ref{eq.rh}) for the definition of $R_H$.
\end{thebibliography}
%===================================================================

\end{document}